\documentstyle [12pt]{article}
\newcommand{\be}{\begin{equation}}
\newcommand{\ee}{\end{equation}}
\newcommand{\ra}{\rightarrow}
\newcommand{\gsim}{\stackrel{>}{\sim}}
\begin{document}
\begin{titlepage}
\begin{flushright}
 IFUP-TH 3/93

 hep-th/9301067

\end{flushright}

\vspace{8mm}

\begin{center}

{\Large\bf A Generalized Uncertainty Principle\\
\vspace{3mm}
in Quantum Gravity}\\
\vspace{12mm}
{\large Michele Maggiore}

\vspace{3mm}

I.N.F.N. and Dipartimento di Fisica dell'Universit\`{a},\\
piazza Torricelli 2, I-56100 Pisa, Italy.\\
\end{center}

\vspace{4mm}

\begin{quote}
ABSTRACT. We discuss a Gedanken experiment for the measurement of the
area of the apparent horizon of a black hole
in quantum gravity. Using rather general
and model-independent considerations we find a generalized uncertainty
principle which agrees with a similar result obtained in the framework of
string theories. The result indicates that a minimum length of the order of
the Planck length emerges naturally from any quantum theory of gravity, and
that the concept of black hole is not operationally
defined if the mass is smaller than the Planck mass.
\end{quote}
\end{titlepage}
\clearpage

1.  In field theories which do not involve gravitation the
transition from a classical to a quantum description is performed
'superimposing' a quantum structure ({\em e.g.}, commutation relations) to the
classical theory; when we attempt to combine gravitation
and quantum theory, however,
we expect that at  distances of the order of the Planck
length the very notion of space-time might need a
radical revision. Because of this, one cannot exclude a priori
that a proper quantum theory of gravity
might require a modification of basic quantum principles.

In this Letter we examine  the uncertainty principle and suggest that
in quantum gravity it is indeed modified. This result, in itself, is not new;
a generalized uncertainty principle has already been proposed
in the context of string theories in refs. [1-2] through an analysis of
Gedanken string collisions at planckian energies [2-4]
(for a recent review, see~\cite{MC}),  and in~\cite{KPP} through a
renormalization group analysis applied to the string.
 Our new point is that we do not consider strings, but
use only general model-independent
properties of a quantum theory of gravitation:
our main physical ingredient is the Hawking radiation~\cite{Haw}.
The functional
form of the
generalized uncertainty principle that we obtain agrees with the one found
in string theory.

Let us consider a Reissner-Nordstr\"{o}m
 black hole with mass $M$ and charge $Q$ (the generalization of our
considerations to
Kerr-Newman black holes does not present conceptual difficulties). The
apparent horizon is  defined~\cite{HE} as the outer boundary of a
region of closed trapped surfaces; it has spherical topology and,
in Boyer-Lindquist coordinates,
it is located at $r=R_h$,
\be
R_h=GM [1+(1-\frac{Q^2}{GM^2})^{1/2}]\, .
\ee
(We set $c=1$ but write explicitly $\hbar$ and $G$ in the
following). In classical general relativity an observer has no
{\em direct }
access to the apparent horizon: no signal is emitted from the black hole.
If an observer wants to obtain the area of the apparent horizon,
he should  measure the mass and charge of the black
hole from the motion of test particles at infinity, and then resort to the
theory, which predicts the relation (1) for $R_h$ as a function of $M$ and
$Q$; the area of the horizon is then $A=4\pi R_h^2$. Alternatively, one can
perform a scattering experiment and again resort to the theory which
predicts the relation between the measured
cross section and $R_h$. For instance, for ultra-relativistic particles
impinging on a Schwarzschild black hole, general relativity predicts a
capture cross section $\sigma =(27/4)\pi R_h^2$.
Instead, it is not possible to measure the area of
the horizon directly, that is, using an apparatus which records photons
emitted by the horizon itself, and without having to resort to
relationships predicted by the theory.

In this sense, from an operational point of view, in classical
general relativity eq.~(1) must be considered as a definition of
$R_h$, rather than a prediction which can be tested experimentally.

In a quantum theory of gravitation, however,
the emission of Hawking radiation allows an observer at
infinity to receive a signal coming from the apparent horizon,
and therefore to perform  a {\em direct} measurement of its area,
at least at the level of a Gedanken experiment.
 The  radius of the horizon and the mass and charge of the
black hole become three quantities subject to independent
experimental determination.
It then makes sense to ask whether the relation given
in eq.~(1) is satisfied. In the following we examine
whether there is an intrinsic
limitation to the precision  of the experimental determination of $R_h$.

First, let us   better specify the setting of our Gedanken experiment.
In the case of a Schwarzschild black hole ($Q=0$) the
Hawking radiation is  emitted spontaneously.
Instead, an extremal ($Q^2=GM^2$)
Reissner-Nordstr\"{o}m black hole  has
zero temperature. In particular, one can
consider a magnetically (rather than electrically) charged black hole
in a world with no magnetically charged particles. In this case the
black hole cannot discharge through pair creation~\cite{Gib}, and
 it is expected to be stable, both classically and
quantum mechanically.
In order to measure directly the area of its apparent horizon
we can consider an experiment
similar to the Heisenberg microscope (fig.~1): a photon
with wavelength $\lambda$ is sent from infinity and absorbed by the black
hole. After absorption, the black hole has a mass $M+\Delta M$, with
$\Delta M=h/\lambda$, and it is not extremal. One expects
that it will decay back to the
extremal state.\footnote{We ignore the
(classically forbidden) possibility
that the black hole may break apart into smaller extremal black holes.
At the quantum level, this possibility has been  suggested
recently~\cite{HH}.
However the existence of
such processes, if triggered by the incident
photon, would only increase the disturbance induced by the measurement,
and therefore can only strengthen the  uncertainty principle that
we will find.}
The emission spectrum in general will not consist of
thermal Hawking radiation, since
 a thermal description breaks down for near-extremal
black holes~\cite{PSSTW}. It might even consist of a single photon, again
of wavelength $\lambda$. For definiteness, we indeed consider the situation
in which a single photon is emitted. If the black hole
rather decays with emission of
many particles (not necessarily photons, of course),
our considerations apply separately to each of them, and
our conclusions are unchanged.
 At 90 degrees with the incoming photon, a
microscope detects the photon emitted (fig.~1). Repeating the experiment
and recording many photons, we obtain an 'image' of the black hole.
 Togheter with a measurement of the distance $d$ of
the black hole from the microscope, this provides a measurement of $R_h$. The
distance $d$ can  be measured
using again the Hawking radiation. One can (in principle) measure the
direction of propagation of  photons emitted at different angles
and trace them back in
order to locate the position of the center of the black hole.

Which experimental precision can be obtained in this
measurement of the radius of the horizon?
Of course, a photon with wavelength $\lambda$ cannot carry
information on a more detailed scale
than $\lambda$ itself. As in the classical Heisenberg analysis, the
resolving power of the microscope gives a
minimum error, $\Delta x^{(1)}$, on $R_h$,
\be\label{D1}
\Delta x^{(1)}\sim\frac{\lambda}{\sin\theta}\, ,
\ee
where $\theta$ is the angle defined in fig.~1. Since the error in the final
momentum of the black hole is $\Delta p\sim h\sin\theta /\lambda$, this
gives the standard Heisenberg relation, $\Delta x^{(1)}\sim\hbar /\Delta p$.
More precisely, from infinity we are actually observing the projection of
the black hole on a $(x,y)$-plane
parallel to the microscope (see fig.~1);
$\Delta x^{(1)}$ is the error on the radius of the
horizon measured along the $x$-direction and $\Delta p$ the error on the
final black hole momentum along the same direction.

Besides, during the emission process the mass of the black hole
varies from  $M+\Delta M$ to $M$, and
the radius of the
horizon changes accordingly.\footnote{We neglect the variation in angular
momentum in the absorption/emission process. It does not alter the
final result.} At the moment of the measure, that is, when the light
quantum is being emitted by the black hole, the quantity that we are
measuring changes discontinuosly.
It does not make sense to ask whether the information carried by the
outgoing photon refers to the black hole immediately before emission,
or immediately after, or
to something in between.  The corresponding error must be considered as
intrinsic to the measurement.
This gives a second source of error on $R_h$ which, for a Schwarzschild
black hole, is $\Delta x^{(2)}\sim 2G\Delta M$ and for a general
Reissner-Nordstr\"{o}m black hole is
\be\label{RN}
\Delta x^{(2)}\sim G\Delta M +\sqrt{(GM+G\Delta M)^2-GQ^2}-\sqrt{(GM)^2
-GQ^2}
\ge 2G\Delta M\, .
\ee
 Since $\Delta M= h /\lambda$, we get an uncertainty
\be\label{D2}
\Delta x^{(2)}\sim\frac{ \hbar G}{\lambda}=\frac{L^2_{\rm Pl}}{\lambda}\, ,
\ee
where $L_{\rm Pl}$ is the Planck length.
Note that $\Delta x^{(2)}$ is only a lower bound on the uncertainty,
and only a Schwarzschild black hole can saturate it. For instance, for an
extremally charged black hole with $M\gg\Delta M$, eq.~(\ref{RN}) gives
$\Delta x^{(2)}\simeq G\sqrt{2M\Delta M}\gg 2G\Delta M$.

 If we combine $\Delta x^{(1)}$ and
$\Delta x^{(2)}$ linearly (this point will be discussed in more detail
below) and use the trivial inequality
\be\label{theta}
\frac{\lambda}{\sin\theta}\ge\lambda\, ,
\ee
we get
\be\label{D0}
\Delta x\gsim\lambda +{\rm const.}\, \frac{L_{\rm Pl}^2}{\lambda}\, .
\ee
In our approach the relative numerical constant between the two terms
cannot be predicted. Indeed, in order to write a relation, like
eq.~(\ref{D0}), which does not depend on the features of the apparatus,
we have eliminated $\theta$ using  the trivial inequality, eq.~(\ref{theta});
the price we pay is  that we cannot consistently estimate the value of
this constant (more on this below).

 Eq.~(\ref{D0}) implies that
there exists a minimum error $(\Delta x)|_{\rm min}\sim {\rm const.}
L_{\rm Pl}$.
Therefore the statement that the radius of the horizon is equal to
the right hand side of eq.~(1) has no operational meaning if we aim at a
precision better that $L_{\rm Pl}$.

It is also suggestive to write a lower bound on $\Delta x$ in terms of
$\Delta p\sim h\sin\theta /\lambda$. In this case, $\Delta x^{(1)}
\sim\hbar /\Delta p$. As for $\Delta x^{(2)}$, remember that
eq.~(\ref{D2}) is due to the discontinuous change of the horizon during the
measurement, and it is not directly related to the uncertainty in
the (x-component) of the black hole momentum after the measure; in fact, it
is present even when the latter is zero (that is, in the limit $\theta\ra
0$). However, we can use the trivial
inequality  $\Delta p /\sin\theta\ge\Delta p$ to obtain
\be\label{D}
\Delta x\gsim \frac{\hbar}{\Delta p}+ c\, G\Delta p\, ,
\ee
again with a relative constant $c$
which our model-independent arguments cannot predict.
\vspace{5mm}

2. It is natural to investigate whether the relation given
in eq.~(\ref{D}), which has
been obtained considering only a very specific measurement, has a more general
validity in quantum gravity. A definite answer to this question presumably
cannot be given using only the rather general
and model-independent arguments that we have
presented, and one should consider the problem within
a specific theory~--{\em e.g.}, string theory. We see two natural
interpretations of eq.~(\ref{D}). One, more restrictive, is that it
expresses nothing more than the fact that, because of quantum fluctuations,
the concept of horizon in
quantum gravity becomes uncertain, and is not defined on a scale smaller
than $L_{\rm Pl}$. This fact, in itself, is not very surprising: it
has already been noted (see {\em e.g.} \cite{CPW}) that if
$M\sim M_{\rm Pl}$
(where $M_{\rm Pl}$ is the Planck mass) the Compton radius of the black
hole approaches its
Schwarzschild radius, and quantum fluctuations in the black hole position
affects the definition of the horizon. Our result, however, is somewhat
stronger; we have shown that the concept of horizon  is not defined,
operationally, at scales smaller than
$ L_{\rm Pl}$; this implies that it does not make
sense to talk of "black holes" with $M< M_{\rm Pl}$, nor to expect that,
for $M\ll M_{\rm Pl}$ (and therefore for elementary particles) some sort of
gravitational substructure, suitably modified by quantum effects,
 exists at scales of order of the Schwarzschild radius. It is more
appropriate to consider the Compton radius $\hbar/M$ and the
Schwarzschild radius $2GM$ as mutually exclusive attributes of an object of
mass $M$, which make sense only in the limits $M\ll M_{\rm Pl}$
and $M\gg M_{\rm Pl}$, respectively.

A second possible  interpretation
of the result that we have found is that it is an
example of a more general situation, and that eq.~(\ref{D}) is indeed
a generalized uncertainty principle which governs all measurement processes
in quantum gravity. This possibility seems to us particularly
attractive, for at least three reasons: 1) it implies that a minimal
observable length emerges naturally from a quantum theory
of gravity. 2) Eq.~(\ref{D}) is
independent of the mass and charge of the black hole and, so to say, has
lost any memory of the particular measurement which was analyzed. We have
seen that in general  $\Delta x^{(2)}$  depends on the mass and charge of
the black hole, see eq.~(\ref{RN}); it is non trivial that in the
$(M,Q)$-plane it has a minimum, non zero, value.
3) Eq.~(\ref{D}) agrees with the
result found in string theory~\cite{GV,ACV1,KPP}.
Notice that, while in
string theory the linear dependence of $\Delta x^{(2)}$  on
$\Delta p$ comes out after a rather technical analysis, in our picture it
is a simple consequence of the linear dependence of the Schwarzschild
radius on the mass. The string theory analysis, on the
other hand, can also predict
(the order of  magnitude of) the constant $c$ in eq.~(\ref{D}), and the
result is $c\sim\alpha '$,
where $\alpha '$ is the string tension (in Planck units).
 It is clear that such a result
cannot be obtained with the general, model-independent,
 considerations that we have presented; from
this point of view, it is welcome that eqs.~(\ref{D0},\ref{D}) could
only be obtained using  the inequality $\sin\theta\le 1$ which, on the one
hand, allows us to get rid of the dependence on the apparatus, but on the
other hand forbids us to give a consistent estimate of the numerical
constant in eqs.~(\ref{D0},\ref{D}); otherwise, it is clear that we
would have necessarily obtained a result of order one. We can
state, however, that this constant is larger than a number of order one
(and hence non-zero), again  in agreement with the string prediction.

Notice further that the pointlike limit is obtained in string theory
letting $\alpha '\ra 0$. In this limit the term linear in $\Delta p$, in
the generalized uncertainty principle predicted from strings, disappears.
Since however we have found, on rather general grounds,
that this term is indeed present, one could take our result as an
{\em indication} of the fact that it is not possible to construct a consistent
quantum theory of gravity with pointlike objects. The same result is
suggested by the emergence of a minimum measurable distance.

In eqs.~(\ref{D0},\ref{D}), for the sake of definiteness, we have
combined $\Delta x^{(1)}$ and $\Delta x^{(2)}$ linearly. More in general,
it is suggestive to introduce a function $R_*(M)$, which
in the proper limits reduces respectively to the Compton radius
$R_C=\hbar /M$
and to the Schwarzschild radius $R_S=2GM$
\be
   R_*(M)\ra\left\{
  \begin{array}{ll}
   \hbar/M     & \mbox{if $M\ll M_{\rm Pl}$}\\
    2GM        & \mbox{if $M\gg M_{\rm Pl}$}
  \end{array}
\right.
\ee
and which reaches a minimum value
$(R_*)|_{\rm min}\sim {\rm const.}\, L_{\rm Pl}$
when $M\sim {\rm const.}\, M_{\rm Pl}$. Loosely speaking,
$R_*$ interpolates between the elementary particle regime
($M\ll M_{\rm Pl})$, where an object
with mass $M$ has a characteristic length $\sim R_C$ and
the black hole regime
($M\gg M_{\rm Pl})$, with characteristic length $\sim R_S$. It is tempting
to interprete this function as the typical length associated with an object
of mass $M$. In particular, this would illustrate the mutually exclusive
nature of $R_C$ and $R_S$ mentioned above; these quantities would
correspond to different limits of the same function; so that it does not
make sense to talk of the Compton radius of a black hole, or of the
Schwarzschild radius of an elementary particle.

In terms of this function, the generalized
uncertainty principle reads
\be\label{R*}
\Delta x\ge R_*(\Delta p)\, .
\ee
Of course, both a confirmation of
the interpretation of $R_*$ and its precise form
can only be obtained by a full quantum theory of gravity.

\vspace{2mm}

I thank A.~Di Giacomo, K.~Konishi, M.~Mintchev and G.~Paffuti for useful
comments.

\clearpage
\begin{center}
{\bf FIGURE CAPTION}
\end{center}
Fig. 1: The Heisenberg microscope experiment described in the text. A
photon moving along the positive $x$-axis is absorbed by
an extremal  black hole; a
microscope detects the induced Hawking radiation, at a distance $d$ along
the $z$-axis; $\theta$ is the angular opening of the microscope. The
projection of the black hole onto the $(x,y)$-plane is a circle whose
radius is measured in the experiment.


\begin{thebibliography}{999}
\newcommand{\pl}{{ Phys. Lett.}\ }
\newcommand{\prl}{{Phys. Rev. Lett.}\ }
\newcommand{\pr}{{ Phys. Rev.}\ }
\newcommand{\np}{{ Nucl. Phys.}\ }
\newcommand{\jl}{{ JEPT Lett.}\ }
\newcommand{\js}{{ Sov. Phys. JEPT }\ }
\newcommand{\bb}{\bibitem}
\bb{GV} G. Veneziano, Europhys. Lett. 2 (1986) 199; Proc. of Texas
Superstring Workshop (1989).

D. Gross, Proc. of ICHEP, Munich (1988).
\bb{ACV1} D. Amati, M. Ciafaloni and G.~Veneziano, \pl B216 (1989) 41.
\bb{ACV2} D. Amati, M. Ciafaloni and G.~Veneziano, \pl B197 (1987) 81;
Int.~J.~Mod. Phys. A3 (1988) 1615;
\np B347 (1990) 530.
\bb{GM} D.J. Gross and P.F. Mende, \pl B197 (1987) 129; \np B303 (1988) 407.
\bb{MC} M.~Ciafaloni, {\em 'Planckian Scattering beyond the Eikonal
Approximation', } preprint DFF 172/9/'92 (1992).
\bb{KPP} K. Konishi, G. Paffuti and P.~Provero, \pl B234 (1990) 276.
\bb{Haw} S.W. Hawking, Comm. Math. Phys. 43 (1975) 199.
\bb{HE} S.W. Hawking and G.F.R. Ellis, {\em The large scale structure of
space-time},  Cambridge University Press, (1973) Cambridge.
\bb{Gib}  G.W. Gibbons, Comm. Math. Phys. 44 (1975) 245.
\bb{HH} D.~Brill, \pr D46 (1992) 1560.

J.H Horne and G.T. Horowitz, {\em "Black Holes Coupled to a Massive
Dilaton"}, preprint UCSBTH-92-17, hep-th/9210012.
\bb{PSSTW} J. Preskill, P. Schwarz, A. Shapere, S.~Trivedi and
F.~Wilczek, Mod. Phys. Lett. A6 (1991) 2353.
\bb{CPW} S.~Coleman, J.~Preskill and F.~Wilczek, \np B378 (1992) 175.
\end{thebibliography}
\end{document}